# Orientation-Dependent β-Ga$_2$O$_3$ Heterojunction Diode with Atomic Layer Deposition (ALD) Grown NiO

Yizheng Liu[1,a)], Shane M.W. Witsell[2], John F. Conley Jr.[2], and Sriram Krishnamoorthy[1]

[1]Department of Materials, University of California Santa Barbara, Santa Barbara CA 93106, USA

[2]School of Electrical Engineering and Computer Science, Oregon State University, Corvallis OR 97331, USA

a) Author(s) to whom correspondence should be addressed. Electronic mail: yizhengliu@ucsb.edu

**Abstract**: This work reports the demonstration of ALD-deposited NiO/β-Ga$_2$O$_3$ heterojunction diodes (HJDs) on low doped drift layer and highly doped (001) & (100) n$^+$ substrates with experimental observation of a parallel-plane junction electric field as high as 7.5 MV/cm, revealing a crystal orientation dependence in β-Ga$_2$O$_3$. We use a novel metalorganic precursor bis(1,4-di-tert-butyl-1,3-diazadienyl) (nickel Ni($^{tBu2}$DAD)$_2$) with ozone (O$_3$) to deposit NiO. The NiO/β-Ga$_2$O$_3$ HJD on 7.7 μm-thick HVPE-grown drift region exhibited an on-state current density of ~20 A/cm$^2$ at 5 V, ~10$^{-8}$ A/cm$^2$ reverse leakage at low reverse bias(-5 V), and a rectifying ratio(J$_{on}$/J$_{off}$) of ~10$^9$. The HJD broke down at ~2.2 kV reverse bias, corresponding to a ~3.4 MV/cm parallel-plane junction electric field, with a noise floor reverse leakage (10$^{-8}$~10$^{-6}$ A/cm$^2$, nA) at 80% of the device catastrophic breakdown voltage. The NiO/β-Ga$_2$O$_3$ HJDs on n$^+$ (001) & (100) highly-doped substrates exhibited breakdown voltages at 12.5-16.0 V and 28.5-70.5 V, respectively, with extracted critical electric field (E$_C$) at 2.30-2.76 MV/cm, and 4.33-7.50 MV/cm, revealing a substrate crystal orientation dependence on breakdown electric field for β-Ga$_2$O$_3$. The 7.5 MV/cm E$_C$ reported here is one of the highest parallel-plane junction electric fields reported in literature.

Wide/ultra-wide bandgap (UWBG) semiconductors are attractive for medium voltage (1-35 kV) grid-scale and renewable energy power electronic applications, such as electronic vehicle (EVs), electric ships (E-ships), fast charging stations ,etc. Some specific examples are silicon carbide (SiC) insulated-gate bipolar transistors (IGBTs) in DC to AC EV motor drives[1], and gallium nitride (GaN) high-electron mobility transistors (HEMTs)[2] in smartphone chargers, where UWBGs are used due to their unique combinations of energy bandgap, electron mobility, and critical electric field strength compared to conventional silicon-based power devices. Among existing UWBG semiconductors, beta gallium oxide (β-Ga$_2$O$_3$) stands out as one of the most promising candidates for extreme voltage, and harsh environment power devices[3] due to its large bandgap (~4.8 eV), shallow hydrogenic dopants (E$_D$~20-30 meV)[4,5], and high critical electric field[6] (E$_C$~8 MV/cm) compared to its counterparts (SiC, GaN, AlN, diamond)[5], along with the availability of large area melt-grown bulk substrates. Because of the lack of reliable p-type homojunctions in β-Ga$_2$O$_3$, high-voltage junctions are often accomplished by using either high dielectric constant Schottky reduced surface field (RESURF) trench diode structures[7–10] on high quality/thick drift regions deposited using halide vapor phase epitaxy (HVPE), metal organic chemical vapor deposition (MOCVD)[11,12], or molecular beam epitaxy (MBE)[13–15], or heterojunctions from sputtered p-type nickel oxide (NiO$_x$)[16–18]. The advantage of UWBG semiconductors in reduced conduction loss is captured by the unipolar figure of merit (FOM, $V_{BR}^2/R_{on,sp} \sim \mu \varepsilon_s E_C^3$), which scales linearly with carrier mobility ($\mu$) and semiconductor relative permittivity ($\varepsilon_s$), and in cubic power with $E_C$.

Operating at higher electric fields is therefore critical for fully realizing the advantages offered by UWBG semiconductors. Although some of the best unipolar Schottky/metal-insulator-semiconductor (MIS) junctions using low conduction band offset materials, such as titanium oxide (TiO$_2$), in β-Ga$_2$O$_3$ can sustain an electric field of 4~5.7 MV/cm[9,19–21] with fast switching performance, larger breakdown field under very high voltage (~10 kV) conditions are often more readily accomplished in P-N junction devices with higher built-in potentials[17,18]. In most of the β-Ga$_2$O$_3$ heterojunction devices reported in literature, NiO$_x$ was deposited via sputtering. As the P-N junction interface may be susceptible to energetic ion damage during the sputtering process, which could lead to premature device breakdown due to inferior junction quality, this deposition method potentially compromises its ability to access higher electric fields.

In this work, we present NiO/β-Ga$_2$O$_3$ heterojunction diodes (HJDs) on low doped (001) HVPE-grown drift layer and n$^+$ highly doped (001) & (100) β-Ga$_2$O$_3$ with NiO grown by atomic layer deposition (ALD) using a novel metalorganic precursor bis(1,4-di-tert-butyl-1,3-diazadienyl) (nickel Ni($^{tBu2}$DAD)$_2$) and ozone (O$_3$). The ALD-grown NiO layer is then capped with p$^{++}$ sputtered NiO$_x$ contact layer and a Ni ohmic anode *in-situ* to preserve the P/N junction quality. The 7.7 μm-thick ($N_D-N_A \sim 9.01 \times 10^{15}$ cm$^{-3}$) HVPE (001)/NiO HJD showed an on-state current density of ~20 A/cm$^2$ at 5 V, ~10$^{-8}$ A/cm$^2$ reverse leakage at low reverse bias(-5 V), and a rectifying ratio($J_{on}/J_{off}$) of ~10$^9$. The n$^+$ (001) & (100) HJDs exhibited breakdown voltages at 12.5-16.0 V and 28.5-70.5 V, respectively, with extracted critical electric field ($E_C$) at 2.30-2.76 MV/cm, and 4.33-7.50 MV/cm, revealing a crystal orientational dependence on $E_C$ for β-Ga$_2$O$_3$. The 7.5 MV/cm $E_C$ reported here is one of the highest parallel-plane junction electric fields reported among existing β-Ga$_2$O$_3$-based junctions.

Atomic layer deposition (ALD) of the NiO layer was performed using a Picosun SUNALE R-150B reactor held at 200 °C. The films were deposited at ~10 Torr using alternating N$_2$-purge-separated pulses of bis(1,4-di-tert-butyl-1,3-diazadienyl) (nickel Ni($^{tBu2}$DAD)$_2$) precursor and O$_3$. The O$_3$ was generated in an IN-USA AC-2025 ozone generator operated at 50% power, yielding an expected mixture of 10% O$_3$/O$_2$. The powdered Ni($^{tBu2}$DAD)$_2$ was heated to 150 °C to ensure vaporization and delivery by a Picosun PicoSolid booster source. Film thickness was measured using a Film Sense FS1000 multi-wavelength ellipsometer and Cauchy modeling. Growth per cycle on a Si witness wafer was calculated to be ~0.12 nm/cycle which matches previous reports of this process[22].

ALD NiO was deposited on 1-inch×1-inch coupons of p-type Si ($\rho$=0.016 $\Omega$/cm) and n-type Si ($\rho$=0.012 $\Omega$/cm) as well as beta gallium oxide ($\beta$-$Ga_2O_3$), as shown in **Fig1.(a)**. Prior to ALD, all sample coupons were submerged into a bath of buffered oxide etch for 1 min to strip any native oxide. During ALD, sample coupons were placed into the ALD reactor atop a 6-in Si carrier wafer. Following ALD, the Si samples were moved to an in-house e-beam evaporator to deposit 50 nm of Al top contact through a shadow mask of 250-μm dia. circular contacts. Contact with the bottom electrode was made through scribing the surface with a diamond scribe and soldering on an indium contact point. The ground connection was set on the opposite edge of the sample from the tested devices, and the indium contact was electrically equivalent to a backside Al blanket coat contact. Current density verses electric field (J-V) measurements were then taken on these Al/NiO/Si metal-oxide-semiconductor (MOS) hetero devices using an Agilent 4156C semiconductor parameter analyzer probe station in a dark box. All electrical characterization was completed at room temperature.

Previous reports by Holden et al[22], demonstrated the p-type nature of ALD deposited NiO. J-V sweeps of these devices in **Fig1.(b)** confirm the p-type nature of NiO (150 cycles, ~18 nm) as the NiO/n-Si samples match a simple P-N junction current response model in both the positive and negative bias, while the NiO/p-Si sample results in a simple threshold response in both the positive and negative bias directions. 165 cycles of ALD NiO were deposited on $\beta$-$Ga_2O_3$ and the thickness was measured to be ~20.3 nm via an Si witness wafer, as shown in **Fig1.(c)**.

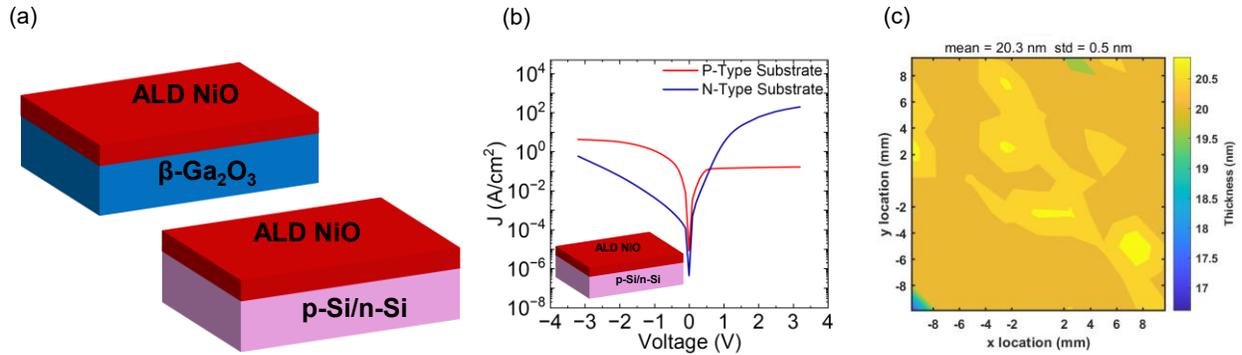

FIG. 1.(a) Schematics of ALD NiO deposited on p-type and n type Si and $\beta$-$Ga_2O_3$. (b) Semi-log-scale J-V characteristics on both p-type and n-type Si. (c) Thickness mapping of ALD deposited NiO on witness wafer via ellipsometry.

Following this ALD NiO deposition, a photoresist lift-off mask was patterned using optical lithography on the (001) Si-doped HVPE $\beta$-$Ga_2O_3$ to prepare a bilayer deposition of conductive $NiO_x$ and *in-situ* Ni Ohmic capping via radio-frequency (RF) magnetron reactive sputtering. During the sputter deposition, a ~30 nm-thick relatively conductive ($R_{sheet}$~14 k$\Omega$/□, $N_a$>>1×10$^{18}$ cm$^{-3}$) $NiO_x$ was deposited on the ALD NiO layer under an oxygen-rich condition (Ar/$O_2$~10/8 sccm) at 3.5 mTorr chamber pressure and 150 W RF power using a 99.999% (5N)- pure metallic nickel (Ni) target[23]. Following the NiO$_x$ deposition, a ~30 nm Ni Ohmic anode cap was continuously sputtered *in-situ* under pure argon (Ar) condition (Ar: 25 sccm) at 5 mTorr and 150 W RF power[23] to minimize the extent of ambient exposure of NiO$_x$. After the sputtered Ni capping, the sample was immediately transferred to electron-beam evaporation for Au/Ni (30/50 nm) stack deposition. The 50-nm Ni served as a hard mask for subsequent self-aligned plasma etching. The NiO$_x$/metal stack was lifted off in a heated n-methyl pyrrolidone (NMP) solution. The exposed ALD NiO and β-Ga$_2$O$_3$ stack were later dry-etched ~1 μm below the NiO/β-Ga$_2$O$_3$ heterojunction interface under inductively coupled BCl$_3$/Ar plasma at 200 W to form rounded edge corners[7,23] for electric field crowding mitigation. To conclude the device fabrication, a backside Ti/Au (80/200 nm) Ohmic cathode stack as deposited on β-Ga$_2$O$_3$ via e-beam evaporation. Details of the device fabrication were illustrated in **Fig.S1** in supplementary materials. Similar fabrication steps were also applied for sputtered NiO$_x$/ALD NiO on (001) and (100) n$^+$ β-Ga$_2$O$_3$ substrates.

Cross-sectional schematics of devices fabricated on HVPE β-Ga$_2$O$_3$ and n$^+$ substrates along with corresponding electric field profiles are shown in **Fig.2**.

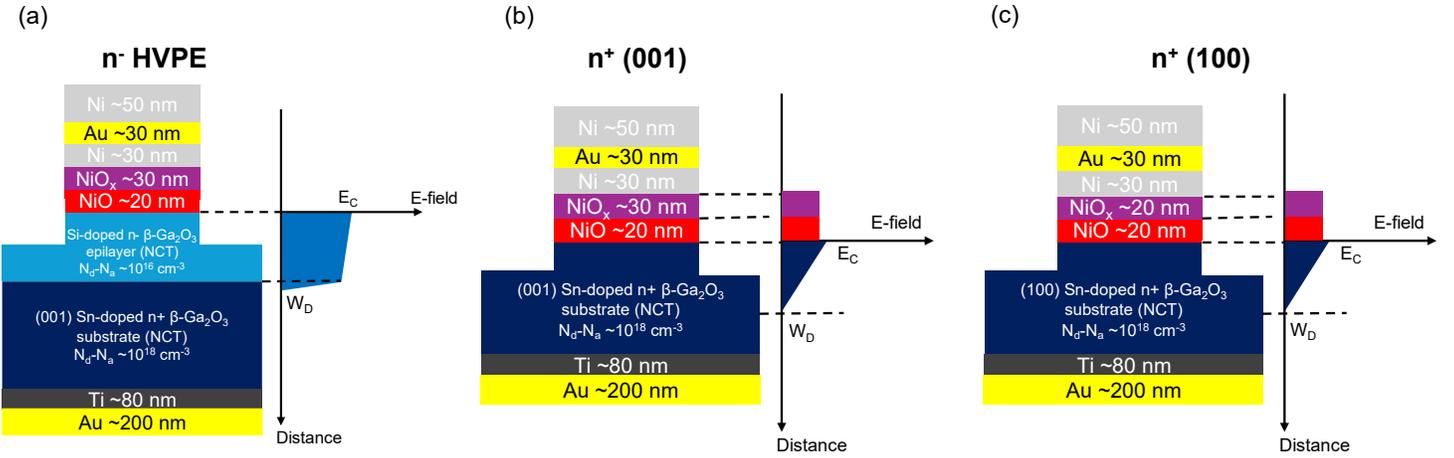

FIG. 2. Device schematics and electric field models of NiO/β-Ga$_2$O$_3$ HJD on (a) (001) HVPE β-Ga$_2$O$_3$, (b) n$^+$ (001) β-Ga$_2$O$_3$ substrate, and (c) n$^+$ (100) β-Ga$_2$O$_3$ substrate.

The punch-through electric field profile of NiO/β-Ga$_2$O$_3$ HJDs beyond ~400 V was confirmed via high-voltage capacitance-voltage measurements at 1 MHz, as shown in **Fig.3(a)**. The HVPE layer was fully depleted above 400 V reverse bias where measured capacitance remained flat and no longer changed as a function of applied reverse bias. Moreover, the built-in potential ($V_{bi}$) of the HVPE HJD was extracted to be ~2 V from the $1/C^2$ vs. voltage characteristics shown in **Fig.3(a)** inset. The average HVPE drift layer's apparent doping density was extracted to be $9.01 \times 10^{15}$ cm$^{-3}$ with a corresponding thickness of ~7.7 μm using a relative permittivity of 12.4 for (001) β-Ga$_2$O$_3$[24], as shown in **Fig.3(b)**.

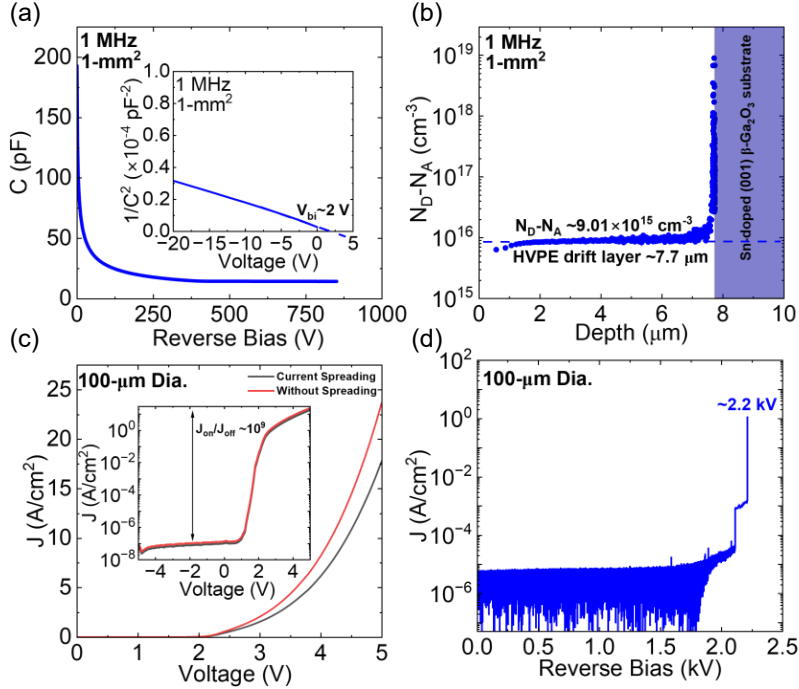

FIG. 3. (a) High-voltage C-V characteristics on 1-mm² area pad at 1 MHz with inset showing the built-in potential of NiO/HVPE β-Ga$_2$O$_3$ HJD. (b) Extracted apparent charge density vs. depth profile of the plasma-etched NiO/HVPE β-Ga$_2$O$_3$ vertical HJD. (c) J-V characteristics of NiO/HVPE β-Ga$_2$O$_3$ HJD on 100-μm dia. device with/without current spreading effect with semi-log scale inset. (d) Breakdown characteristics of 100-μm dia. NiO/HVPE β-Ga$_2$O$_3$ HJD.

The forward linear J-V characteristics of the HVPE HJD on the 100-μm dia.device are shown in **Fig.3(c)** by considering a 45°-angle current spreading[23,25] model into the drift region and a one without spreading effect. The relatively low forward on-current density can be possibly attributed to the resistive nature of the ALD-grown NiO. The rectifying ratio of HVPE HJD is ~10$^9$ with a reverse leakage of ~10$^{-8}$ A/cm² at -5 V as shown in **Fig.3(c)** inset. The reverse leakage and breakdown characteristics of the HVPE HJD (submerged in silicone oil) for a 100-μm dia.device are shown in **Fig.3(d)**, showcasing a breakdown voltage at ~2.2 kV with a noise-floor reverse leakage current density (10$^{-8}$~10$^{-6}$ A/cm², nA) at 80% of the device's catastrophic breakdown voltage.

The parallel-plane junction electric field (E$_C$) of the HVPE HJD on 100-μm device was calculated using Eqn. (1)[1].

$$BV_{PT} = E_C W_{D,PT} - \frac{qN_{D,PT}W_{D,PT}^2}{2\varepsilon_s} \qquad (1)$$

For a breakdown voltage (BV$_{PT}$) at ~2.2 kV, punch-through depletion thickness (W$_{D,PT}$) of ~7.7 μm, an HVPE drift region apparent doping density (N$_D$) of ~9.01×10$^{15}$ cm⁻³, and a relative vacuum permittivity (ε$_s$) of 12.4ε$_0$[24] for (001)-orientated β-Ga$_2$O$_3$, the parallel-plane junction critical electric field of NiO/HVPE β-Ga$_2$O$_3$ vertical HJD was accurately extracted to be E$_C$ ~3.4 MV/cm without assuming the nominal doping density and drift layer thickness.

Similar E$_C$ analysis was also conducted on both n$^+$ (100) and (001) β-Ga$_2$O$_3$ substrates. The doping profile of each substrate is shown in **Fig.4(a)** with ~3.38×10$^{18}$ cm$^{-3}$ and ~7.2×10$^{18}$ cm$^{-3}$ for (100) and (001) n$^+$ β-Ga$_2$O$_3$, respectively. The inset of **Fig.4(a)** also revealed the V$_{bi}$ of HJDs on both substrates to be ~ 2 V.

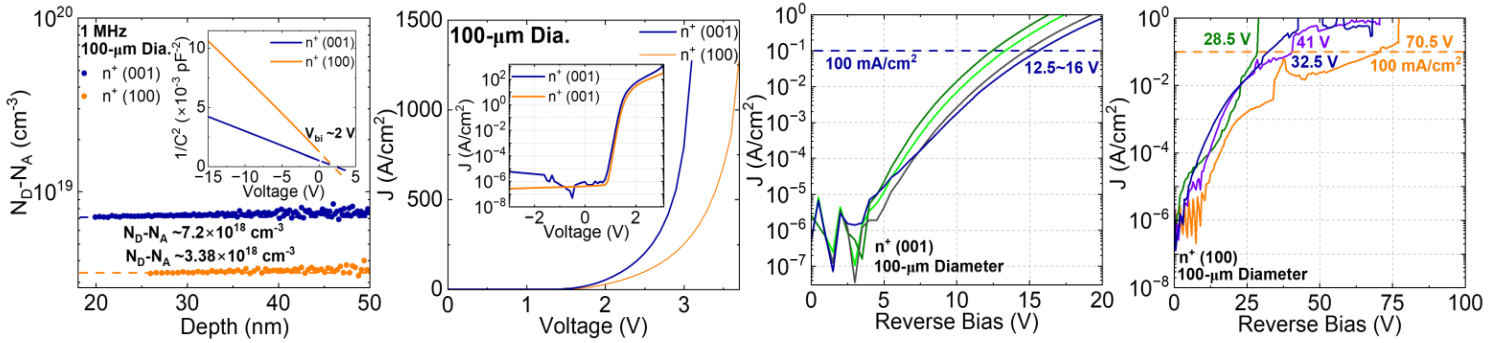

FIG.4. (a) Extracted apparent charge density vs. depth profile of the plasma-etched NiO/n$^+$ (100) &(001) β-Ga$_2$O$_3$ vertical HJDs. (b) J-V characteristics of 100-μm dia. NiO/n$^+$ (100) & (001) β-Ga$_2$O$_3$ vertical HJDs with semi-log scale inset. (c) Breakdown characteristics of 100-μm dia. NiO/(001) n$^+$ β-Ga$_2$O$_3$ HJD with 100 mA/cm$^2$ as the breakdown current density in Fluorinert liquid. (d) Breakdown characteristics of 100-μm dia. NiO/(100) n$^+$ β-Ga$_2$O$_3$ HJDs with 100 mA/cm$^2$ as the breakdown current density in Fluorinert liquid.

The forward linear J-V characteristics of 100-μm dia. HJDs on highly doped n$^+$ substrates, shown in **Fig.4(b)**, both exhibited a forward current density exceeding 1000 A/cm$^2$ at 4 V due to their elevated charge density compared to the HVPE HJD. The reverse leakage current density of HJDs for both β-Ga$_2$O$_3$-crystal-orientations (**Fig.4(b)** inset) inset is between 10$^{-8}$~10$^{-6}$ A/cm$^2$ at -4 V with a rectifying ratio of ~10$^9$ to 10$^{11}$. The breakdown characteristics of the n$^+$ (001) HJDs are shown in **Fig.4(c)**, revealing a breakdown voltage at 12.5-16.0V when defining a current density of 100 mA/cm$^2$ as the point for device breakdown. Similarly, the n$^+$ (100) HJDs broke down in the range of 28.5-41.0 V with the highest breakdown voltage at 70.5 V, as shown in **Fig.4(d)**. Devices on both highly doped substrates were submerged in Fluorinert liquid (FC-40) during breakdown measurements.

As both (100) and (001) substrates were highly doped, a triangular non-punch-through electric field profile into the n$^+$ β-Ga$_2$O$_3$ was assumed, as shown in **Fig.2(b)&(c)**. To extract the lower-bound value of the parallel-plane critical electric field at breakdown, we assumed that both ALD NiO and sputtered NiO$_x$ were fully depleted with a rectangular field profile within both layers. Therefore, the total voltage at device breakdown can be expressed as

$$t_{NiO}^{total} \times \frac{q(N_D-N_A)W_n}{\varepsilon_{s(NiO_x)}} + \frac{1}{2} \times \frac{q(N_D-N_A)W_n^2}{\varepsilon_{s(\beta-Ga_2O_3)}} = V_{100\ mA/cm^2} + V_{bi} \qquad (2)$$

Assuming a relative permittivity of $11.9^{18}$ for NiO$_x$ ($\varepsilon_{S(NiO_x)}$) and $12.4^{24}$ for (001) β-Ga$_2$O$_3$ ($\varepsilon_{S(\beta-Ga_2O_3)}$) with a total combined of NiO thickness ($t_{NiO}^{total}$) of 50 nm ($t_{NiO}^{ALD}+t_{NiO}^{sputtered}$), a net apparent doping density of 7.2×10$^{18}$ cm$^{-3}$ (N$_D$-N$_A$), and a built-in potential (V$_{bi}$) of 2 V, the depletion thickness (W$_n$) at reverse bias breakdown into the n$^+$ (001) β-Ga$_2$O$_3$ was calculated to be 21.9~26.3 nm, which corresponds to a parallel-plane junction electric field of 2.30~2.76 MV/cm in n$^+$ (001) β-Ga$_2$O$_3$. Similarly, the lower-bound value of E$_C$ of n$^+$ (100) β-Ga$_2$O$_3$ was extracted to be 4.33~7.50 MV/cm using a $\varepsilon_{S((100)\beta-Ga_2O_3)} = 10.2^{24}$, showcasing one of the highest E$_C$ values reported in β-Ga$_2$O$_3$ devices while indicating a strong E$_C$ dependence on β-Ga$_2$O$_3$ crystal orientation, as described in **Fig.5(a)**.

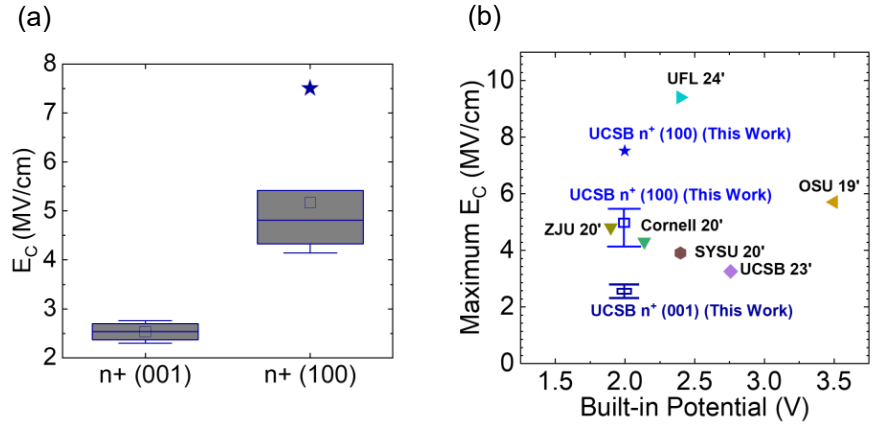

FIG.5. (a) E$_C$ distribution at breakdown for both n$^+$ (001) and (100) β-Ga$_2$O$_3$ substrates. (b) Benchmarking of E$_C$ for β-Ga$_2$O$_3$ vertical power devices[9,16,17,19,23,26].

Moreover, in the benchmarking of E$_C$ vs. device built-in potential, as shown in **Fig.5(b)**, the ALD NiO/n$^+$ (100) β-Ga$_2$O$_3$ HJD exhibited one of the highest E$_C$'s at ~2 V built-in potential among existing β-Ga$_2$O$_3$ vertical power rectifiers[9,16,17,19,23,26].

In summary, this work demonstrated β-Ga$_2$O$_3$ heterojunction diodes using ALD grown NiO on low doped (001) HVPE drift region substrate with ~2.2 kV breakdown voltages and noise-floor reverse leakage current densities (10$^{-8}$~10$^{-6}$ A/cm$^2$, nA) at 80% of the device's catastrophic breakdown voltage, exhibiting a parallel-plane junction electric field of ~3.4 MV/cm. For n$^+$ (001) and (100) β-Ga$_2$O$_3$ HJDs, a parallel-plane junction electric field of 2.30-2.76 MV/cm and 4.33-7.50 MV/cm was extracted, respectively, revealing an orientational dependence while exhibiting one of the highest experimentally observed critical electric fields (7.5 MV/cm) among β-Ga$_2$O$_3$ power devices.

**ACKNOWLEDGMENTS**

The authors acknowledge funding from the U.S. Department of Energy (DOE) ARPA-E ULTRAFAST program (DE-AR0001824) and Coherent/II-VI Foundation Block Gift Program. A portion of this work was performed at the UCSB Nanofabrication Facility, an open access laboratory. Y. Liu would like to acknowledge the electrical data measurements from D. Lopez and P. Bursin. S.W. and J. Conley. and S. Witsell would like to acknowledge the in-kind support of EMD Electronics (Sheboygan Falls, WI) in supplying the nickel precursor used for ALD as well as Lam Research (Tualatin) for gift funding. A portion of the work was also conducted at the Materials Synthesis and Characterization (MaSC) Center, a National Science Foundation (NSF) National Nanotechnology Coordinated Infrastructure (NNCI) Northwest Nanotechnology Infrastructure (NWNI) open user facility at Oregon State University supported in part through NSF Grant NNCI-2025489.

**SUPPLEMENTARY MATERIAL**

See the supplementary material for the detailed fabrication schematics of NiO/(001) HVPE β-$Ga_2O_3$ heterojunction diode.

**DATA AVAILABILITY**

The data that supports the findings of this study are available from the corresponding authors upon reasonable request.

**REFERENCES**

[1] B.J. Baliga, *Fundamentals of Power Semiconductor Devices* (Springer Science & Business Media, 2010).

[2] U.K. Mishra, and M. Guidry, "Lateral GaN Devices for Power Applications (from kHz to GHz)," in *Power GaN Devices: Materials, Applications and Reliability*, edited by M. Meneghini, G. Meneghesso, and E. Zanoni, (Springer International Publishing, Cham, 2017), pp. 69–99.

[3] A.J. Green, J. Speck, G. Xing, P. Moens, F. Allerstam, K. Gumaelius, T. Neyer, A. Arias-Purdue, V. Mehrotra, A. Kuramata, K. Sasaki, S. Watanabe, K. Koshi, J. Blevins, O. Bierwagen, S. Krishnamoorthy, K. Leedy, A.R. Arehart, A.T. Neal, S. Mou, S.A. Ringel, A. Kumar, A. Sharma, K. Ghosh, U. Singisetti, W. Li, K. Chabak, K. Liddy, A. Islam, S. Rajan, S. Graham, S. Choi, Z. Cheng, and M. Higashiwaki, "β-Gallium oxide power electronics," APL Materials **10**(2), 029201 (2022).

[4] A.T. Neal, S. Mou, S. Rafique, H. Zhao, E. Ahmadi, J.S. Speck, K.T. Stevens, J.D. Blevins, D.B. Thomson, and N. Moser, "Donors and deep acceptors in β-$Ga_2O_3$," Applied Physics Letters **113**(6), (2018).

[5] Y. Zhang, and J.S. Speck, "Importance of shallow hydrogenic dopants and material purity of ultra-wide bandgap semiconductors for vertical power electron devices," Semiconductor Science and Technology **35**(12), 125018 (2020).

[6] A.J. Green, K.D. Chabak, E.R. Heller, R.C. Fitch, M. Baldini, A. Fiedler, K. Irmscher, G. Wagner, Z. Galazka, and S.E. Tetlak, "3.8-MV/cm Breakdown Strength of MOVPE-Grown Sn-Doped β-$Ga_2O_3$ MOSFETs," IEEE Electron Device Letters **37**(7), 902–905 (2016).

[7] S. Roy, B. Kostroun, J. Cooke, Y. Liu, A. Bhattacharyya, C. Peterson, B. Sensale-Rodriguez, and S. Krishnamoorthy, "Ultra-low reverse leakage in large area kilo-volt class β-$Ga_2O_3$ trench Schottky barrier diode with high-k dielectric RESURF," Applied Physics Letters **123**(24), (2023).

[8] S. Roy, B. Kostroun, Y. Liu, J. Cooke, A. Bhattacharyya, C. Peterson, B. Sensale-Rodriguez, and S. Krishnamoorthy, "Low $Q_CV_F$ 20A/1.4 kV β-$Ga_2O_3$ Vertical Trench High-k RESURF Schottky Barrier Diode with Turn-on Voltage of 0.5 V," IEEE Electron Device Letters, (2024).

[9] Z. Xia, H. Chandrasekar, W. Moore, C. Wang, A.J. Lee, J. McGlone, N.K. Kalarickal, A. Arehart, S. Ringel, F. Yang, and S. Rajan, "Metal/$BaTiO_3$/β-$Ga_2O_3$ dielectric heterojunction diode with 5.7 MV/cm breakdown field," Applied Physics Letters **115**(25), 252104 (2019).

[10] W. Li, Z. Hu, K. Nomoto, R. Jinno, Z. Zhang, T. Tu, K. Sasaki, A. Kuramata, D. Jena, and H.G. Xing, "2.44 kV Ga 2 O 3 vertical trench Schottky barrier diodes with very low reverse leakage current," IEDM Tech," Dig **8**, 1–8 (2018).

[11] A. Bhattacharyya, C. Peterson, K. Chanchaiworawit, S. Roy, Y. Liu, S. Rebollo, and S. Krishnamoorthy, "Over 6 μm thick MOCVD-grown low-background carrier density (1015 cm-3) high-mobility (010) β-$Ga_2O_3$ drift layers," Applied Physics Letters **124**(1), (2024).

[12] C. Peterson, A. Bhattacharyya, K. Chanchaiworawit, R. Kahler, S. Roy, Y. Liu, S. Rebollo, A. Kallistova, T.E. Mates, and S. Krishnamoorthy, "200 $cm^2$/Vs electron mobility and controlled low $10^{15}$ $cm^{-3}$ Si doping in (010) β-$Ga_2O_3$ epitaxial drift layers," Applied Physics Letters **125**(18), (2024).

[13] S. Rebollo, Y. Liu, C. Peterson, S. Krishnamoorthy, and J.S. Speck, "Growth of nitrogen-doped (010) β-$Ga_2O_3$ by plasma-assisted molecular beam epitaxy using an $O_2$/$N_2$ gas mixture," Applied Physics Letters **126**(8), (2025).

[14] A. Mauze, Y. Zhang, T. Itoh, E. Ahmadi, and J.S. Speck, "Sn doping of (010) β-$Ga_2O_3$ films grown by plasma-assisted molecular beam epitaxy," Applied Physics Letters **117**(22), (2020).

[15] T. Itoh, A. Mauze, Y. Zhang, and J.S. Speck, "Epitaxial growth of β-$Ga_2O_3$ on (110) substrate by plasma-assisted molecular beam epitaxy," Applied Physics Letters **117**(15), (2020).

[16] J.-S. Li, H.-H. Wan, C.-C. Chiang, T.J. Yoo, M.-H. Yu, F. Ren, H. Kim, Y.-T. Liao, and S.J. Pearton, "Breakdown up to 13.5 kV in NiO/β-$Ga_2O_3$ vertical heterojunction rectifiers," ECS Journal of Solid State Science and Technology **13**(3), 035003 (2024).

[17] J. Wan, H. Wang, C. Zhang, Y. Li, C. Wang, H. Cheng, J. Li, N. Ren, Q. Guo, and K. Sheng, "3.3 kV-class NiO/β-$Ga_2O_3$ heterojunction diode and its off-state leakage mechanism," Applied Physics Letters **124**(24), (2024).

[18] M. Xiao, B. Wang, J. Spencer, Y. Qin, M. Porter, Y. Ma, Y. Wang, K. Sasaki, M. Tadjer, and Y. Zhang, "NiO junction termination extension for high-voltage (> 3 kV) $Ga_2O_3$ devices," Applied Physics Letters **122**(18), (2023).

[19] D. Saraswat, W. Li, K. Nomoto, D. Jena, and H.G. Xing, "Very High Parallel-Plane Surface Electric Field of 4.3 MV/cm in $Ga_2O_3$ Schottky Barrier Diodes with PtOx Contacts," in *2020 Device Research Conference (DRC)*, (2020), pp. 1–2.

[20] S. Dhara, N.K. Kalarickal, A. Dheenan, C. Joishi, and S. Rajan, "β-Ga$_2$O$_3$ Schottky barrier diodes with 4.1 MV/cm field strength by deep plasma etching field-termination," Applied Physics Letters **121**(20), (2022).

[21] J. Williams, N. Hendricks, W. Wang, A. Adams, J. Piel, D. Dryden, K. Liddy, N. Sepelak, B. Morell, and A. Miesle, "Ni/TiO$_2$/β-Ga$_2$O$_3$ Heterojunction Diodes with NiO Guard Ring Simultaneously Increasing Breakdown Voltage and Reducing Turn-on Voltage," in *2023 Device Research Conference (DRC)*, (IEEE, 2023), pp. 1–2.

[22] K.E.K. Holden, C.L. Dezelah, and J.F. Conley, "Atomic Layer Deposition of Transparent p-Type Semiconducting Nickel Oxide Using Ni($^{t}$Bu$_2$ DAD)$_2$ and Ozone," ACS Appl. Mater. Interfaces **11**(33), 30437–30445 (2019).

[23] Y. Liu, S. Roy, C. Peterson, A. Bhattacharyya, and S. Krishnamoorthy, "Ultra-low reverse leakage NiO$_x$/β-Ga$_2$O$_3$ heterojunction diode achieving breakdown voltage> 3 kV with plasma etch field-termination," AIP Advances **15**(1), (2025).

[24] A. Fiedler, R. Schewski, Z. Galazka, and K. Irmscher, "Static dielectric constant of β-Ga$_2$O$_3$ perpendicular to the principal planes (100),(010), and (001)," ECS Journal of Solid State Science and Technology **8**(7), Q3083 (2019).

[25] S. Roy, A. Bhattacharyya, C. Peterson, and S. Krishnamoorthy, "2.1 kV (001)-β-Ga$_2$O$_3$ vertical Schottky barrier diode with high-k oxide field plate," Applied Physics Letters **122**(15), (2023).

[26] X. Lu, X. Zhou, H. Jiang, K.W. Ng, Z. Chen, Y. Pei, K.M. Lau, and G. Wang, "1-kV Sputtered p-NiO/n-Ga$_2$O$_3$ Heterojunction Diodes With an Ultra-Low Leakage Current Below 1~μA/cm$^2$," IEEE Electron Device Letters **41**(3), 449–452 (2020).